\pdfoutput=1
\pdfpageattr{/Group << /S /Transparency /I true /CS /DeviceRGB>>}
\documentclass[5p,times,merge,number,sort&compress]{elsarticle}
\usepackage[T1]{fontenc}
\usepackage[utf8]{inputenc}
\usepackage[english]{babel}

\usepackage{isotope}
\usepackage{xcolor}
\usepackage{graphicx}
\usepackage[load-configurations=abbreviations]{siunitx}
\usepackage{braket}
\usepackage{mathtools}
\usepackage{ifpdf}
\usepackage{booktabs}

\usepackage{txfonts}

\ifpdf\else\renewcommand{\href}[2]{#2}\fi

\let\rtxappendix\appendix
\usepackage{hyperref}
\let\Hyappendix\appendix
\AtBeginDocument{\renewcommand*{\appendix}{\Hyappendix\rtxappendix}}

\usepackage[capitalise]{cleveref}

\crefname{equation}{}{}
\Crefname{equation}{Equation}{Equations}

\makeatletter
\DeclareUnicodeCharacter{014C}{\@tabacckludge=O} 
\DeclareUnicodeCharacter{014D}{\@tabacckludge=o} 
\DeclareUnicodeCharacter{2212}{--} 
\makeatother

\DeclareSIUnit{\GeV}{\giga\electronvolt}
\DeclareSIUnit{\MeV}{\mega\electronvolt}
\DeclareSIUnit{\keV}{\kilo\electronvolt}
\DeclareSIUnit\MeVc{\MeV\per\text{\ensuremath{c}}}
\DeclareSIUnit{\fm}{\femto\meter}

\newcommand{\partd}[2]{\frac{\partial #1}{\partial #2}}

\newcommand{\comm}[2]{\lbrack #1,#2\rbrack}

\newcommand{\Nmax}{\ensuremath{N_\text{max}}}

\newcommand*{\op}[1]{\boldsymbol{#1}}
\newcommand{\Ham}{\op{H}}
\newcommand{\Tint}{\op{T}_{\mspace{-5mu}\text{int}}}

\newcommand{\NNN}{\ensuremath{NNN}}
\newcommand{\NN}{\ensuremath{NN}}
\newcommand{\YN}{\ensuremath{YN}}
\newcommand{\YNN}{\ensuremath{YNN}}

\allowdisplaybreaks[1]

\begin{document}

\title{Light Neutron-Rich Hypernuclei from the Importance-Truncated No-Core Shell Model}

\author{Roland Wirth}
\ead{roland.wirth@physik.tu-darmstadt.de}
\author{Robert Roth}
\ead{robert.roth@physik.tu-darmstadt.de}
\address{Institut für Kernphysik -- Theoriezentrum, Technische Universität Darmstadt, Schlossgartenstr. 2, 64289 Darmstadt, Germany}

\date{\today}

\begin{abstract}%
We explore the systematics of ground-state and excitation energies in singly-strange hypernuclei throughout the helium and lithium isotopic chains --- from \isotope[5][\Lambda]{He} to \isotope[11][\Lambda]{He} and from \isotope[7][\Lambda]{Li} to \isotope[12][\Lambda]{Li} --- in the \emph{ab initio} no-core shell model with importance truncation. All calculations are based on two- and three-baryon interaction from chiral effective field theory and we employ a similarity renormalization group transformation consistently up to the three-baryon level to improve the model-space convergence.
While the absolute energies of hypernuclear states show a systematic variation with the regulator cutoff of the hyperon-nucleon interaction, the resulting neutron separation energies are very stable and in good agreement with available data for both nucleonic parents and their daughter hypernuclei. We provide predictions for the neutron separation energies and the spectra of neutron-rich hypernuclei that have not yet been observed experimentally. Furthermore, we find that the neutron drip lines in the helium and lithium isotopic chains are not changed by the addition of a hyperon.
\end{abstract}

\maketitle

\section{Introduction}

The exploration of the extremes of nuclear existence is one of the main drivers in low-energy nuclear physics today.
Current and future experimental facilities, like FAIR, FRIB, JLab, J-Parc, or RIBF, strive for more and more neutron-rich nuclei, approaching the neutron drip line.
The structure of nuclei with large neutron excess provides valuable information about less-constrained parts of the nuclear interaction and is a challenge for nuclear theory.
Light neutron-rich nuclei are an ideal testing ground for exploring nuclear interactions at large neutron-to-proton ratios.
At the same time, heavier neutron-rich nuclei play a crucial role in nucleosynthesis processes in astrophysical environments, i.e., the $r$ process responsible for the production of the majority of heavy elements in the universe \cite{Arnould2007}.
Likewise, the strong interaction at the neutron-rich extremes governs the structure and stability of neutron stars \cite{Heiselberg2000, Hebeler2015}.

Strangeness in nuclei has also been a focus of experimental and theoretical activity \cite{Gal2016}.
A recent highlight are the mirror hypernuclei \isotope[4][\Lambda]{H} and \isotope[4][\Lambda]{He}, which exhibit a marked charge-symmetry breaking effect \cite{Yamamoto2015,Gazda2016a,Gazda2016b,Schulz2016}.
Beyond these very light systems, which can be described theoretically with established \emph{ab initio} few-body methods \cite{Nemura2002,Nogga2013}, a multitude of phenomenological models like mean-field \cite{Rufa1987,Mares1989}, Skyrme \cite{Rayet1981,Cugnon2000,Tretyakova1999,*Tretyakova2001}, cluster \cite{Motoba1983,Hiyama2009} or microscopic shell models \cite{Gal1971,*Gal1972,*Gal1978,Millener2008,*Millener2010,*Millener2012} have been used to describe heavier hypernuclei.
Also, quantum Monte Carlo methods have been developed \cite{Lonardoni2013,Lonardoni2014}, which can calculate ground-state energies throughout a large part of the hypernuclear chart but are limited to simplified interactions.
Recently, we presented a powerful \emph{ab initio} method suitable for $p$\nobreakdash-shell hypernuclei: the importance-truncated no-core shell model (IT-NCSM) for hypernuclei \cite{Wirth2014}.
With the IT-NCSM we can compute not only ground, but also excited states including all relevant electromagnetic observables \cite{Maris2014,Calci2016a}.
In order to accelerate the convergence of the IT-NCSM we employ similarity renormalization group (SRG) transformations and we recently extended the SRG to hyperon-nucleon ($\YN$) and induced hyperon-nu\-cle\-on-nucleon ($\YNN$) interactions \cite{Wirth2016}.

In this work, we connect the worlds of neutron-rich nuclei and strangeness.
We explore light neutron-rich hypernuclei and study the impact of the additional hyperon on their structure.
In particular, we consider the helium and lithium isotopic chains and their hypernuclear analogs.
Some of these hypernuclei have been studied in experiment \cite{Davis2005,Saha2005,Nakamura2013,Urciuoli2015}, others can in principle be produced but have not been observed \cite{Agnello2012b}.
Some are not accessible in experiments that produce hypernuclei off stable targets.
However, the possibility of using heavy ion collisions to produce hypernuclei of widely varying mass and proton-neutron asymmetry is actively discussed \cite{ToporPop2010,Botvina2017} so that these hypernuclei may become accessible in the future.

While some hypernuclei from these isotopic chains have been considered in a microscopic shell model \cite{Millener2012,Gal2013}, this work provides the first systematic full \emph{ab initio} treatment with interactions from chiral effective field theory.
Aside from exploring the low-lying spectra of these hypernuclei we investigate whether the binding provided by the hyperon-nucleon interaction shifts the neutron drip line compared to the nonstrange isotopes.

\section{Importance-Truncated No-Core Shell Model}
We compute the hypernuclear spectra using the IT-NCSM for hypernuclei \cite{Wirth2014,HyperNCSMTech} including the $\Lambda$ and $\Sigma$ hyperons explicitly.
We start from a Hamiltonian
\begin{equation}
  \Ham = \Delta\op{M} + \Tint + \op{V}_{\NN} + \op{V}_{\NNN} + \op{V}_{\YN}
\end{equation}
that contains the nucleonic two- and three-body interactions $\op{V}_{\NN}$ and $\op{V}_{\NNN}$, and a hyperon-nucleon interaction $\op{V}_{\YN}$.
The first term $\Delta\op{M}$ is a mass term that accounts for the different rest masses of the $\Lambda$ and $\Sigma$ hyperons; $\Tint$ is the intrinsic kinetic energy.
We use the physical masses of the proton, neutron and the hyperons in the calculation of these terms.

This Hamiltonian contains significant short-range correlations due to short-range repulsions and tensor forces, so that a calculation with this ``bare'' Hamiltonian requires exceedingly large model spaces in order to get converged energies.
We improve the convergence behavior of the Hamiltonian using an SRG transformation that suppresses these correlations and thus reduces the model-space dimensions required for convergence.
The SRG \cite{Wegner1994,Wegner2000,Bogner2007} is a very general family of unitary transformations that transforms the Hamiltonian according to the flow equation
\begin{equation}
  \partd{\Ham_{\alpha}}{\alpha} = \comm{\op{\eta}_{\alpha}}{\Ham_{\alpha}}, \label{eq:srg flow}
\end{equation}
where $\op{\eta}_{\alpha}$ is the anti-Hermitian generator of the transformation and $\alpha$ is the flow parameter.
We adopt the common choice for the generator in nuclear physics \cite{Bogner2007}
\begin{equation}
  \op{\eta}_{\alpha} = m_N^2\comm{\Tint}{\Ham_{\alpha}},
\end{equation}
where the nucleon mass $m_N$ fixes the units of $\alpha$.

The SRG flow induces many-body terms: the commutator on the right-hand side of \cref{eq:srg flow} initially contains up to four-body terms when $\Ham_{\alpha}$ is a two-body operator.
Thus, for any finite flow parameter $\alpha$, the evolved Hamiltonian $\Ham_{\alpha}$ of an $A$\nobreakdash-body system consists of up to $A$\nobreakdash-body terms.
Since using the full $A$\nobreakdash-body operator is computationally not feasible, we need to truncate the evolved Hamiltonian at some lower operator-rank.
We already include initial three-body interactions for the nucleonic part and the induced four-body terms are small for light nuclei \cite{Roth2014}, therefore we keep terms up to the three-body level.
The hyperonic part initially consists only of a two-body interaction.
However, we recently showed \cite{Wirth2016} that the induced hyperon-nucleon-nucleon (\YNN{}) terms are strong and it is vital to include them in the calculation to get reliable energies and spectra.
Thus, we also keep the induced \YNN{} terms and the evolved Hamiltonian that is used in the IT-NCSM is
\begin{equation}
  \Ham_{\alpha} = \Delta\op{M} + \Tint + \tilde{\op{V}}_{\NN,\alpha} + \tilde{\op{V}}_{\NNN,\alpha} + \tilde{\op{V}}_{\YN,\alpha} + \tilde{\op{V}}_{\YNN,\alpha},
\end{equation}
where corrections to the mass term and intrinsic kinetic energy have been absorbed into the interaction terms.
The induced \YNN{} terms are computed by embedding the Hamiltonian into a three-body basis spanned by HO states with respect to relative Jacobi coordinates, performing the SRG evolution, and subtracting from the result the Hamiltonian evolved in two-body space.
Details can be found in Ref.~\cite{YNNTech}.

The no-core shell model (NCSM) is based on an expansion of the many-body wave function in terms of Slater determinants built from harmonic-oscillator single-particle states.
We define a finite model space by limiting the total number of oscillator quanta through
$
  \sum_{i=1}^A 2n_i + l_i \le \Nmax + N_0,
$
where $n_i$ and $l_i$ are the radial and orbital angular momentum quantum numbers of the $i$\textsuperscript{th} particle, and $N_0$ is the number of quanta in the lowest state allowed by the Pauli principle.
Since the \YN{} interaction allows changing particle types through $\Lambda$--$\Sigma$ conversion, we have to include all particle combinations permitted by the given charge, strangeness and isospin projection of the system under consideration.
To solve the many-body problem and obtain the eigenenergies and eigenstates of the Hamiltonian, we construct a matrix representation of $\Ham_{\alpha}$ in the Slater-determinant basis and diagonalize it.
As we increase $\Nmax$ the eigenvalues of the matrix approach those of the Hamiltonian from above (variational principle) and convergence indicates that we have isolated an eigenvalue of the many-body Hamiltonian.

The dimension of the model space increases quickly with $\Nmax$ and $A$ so that calculations even for moderate values of the truncation parameter become computationally difficult.
However, many of the basis states in the model space have only a small overlap with the low-lying eigenstates of the Hamiltonian that we are interested in.
By selecting only relevant basis states for inclusion into the model space via an importance measure derived from perturbation theory, we reduce the dimension of the importance-truncated model space by orders of magnitude compared to the full NCSM model space.
The residual effect of the neglected basis states on observables is accounted for by extrapolating the expectation values to the full model space.
The importance truncation of the model space with subsequent extrapolation constitutes the IT-NCSM, which is explained in detail in Ref.~\cite{HyperNCSMTech}.

\section{Light Neutron-Rich Hypernuclei}
\nocite{Entem2003,Navratil2007,Polinder2006}
\begin{figure}
  \centering
  \includegraphics{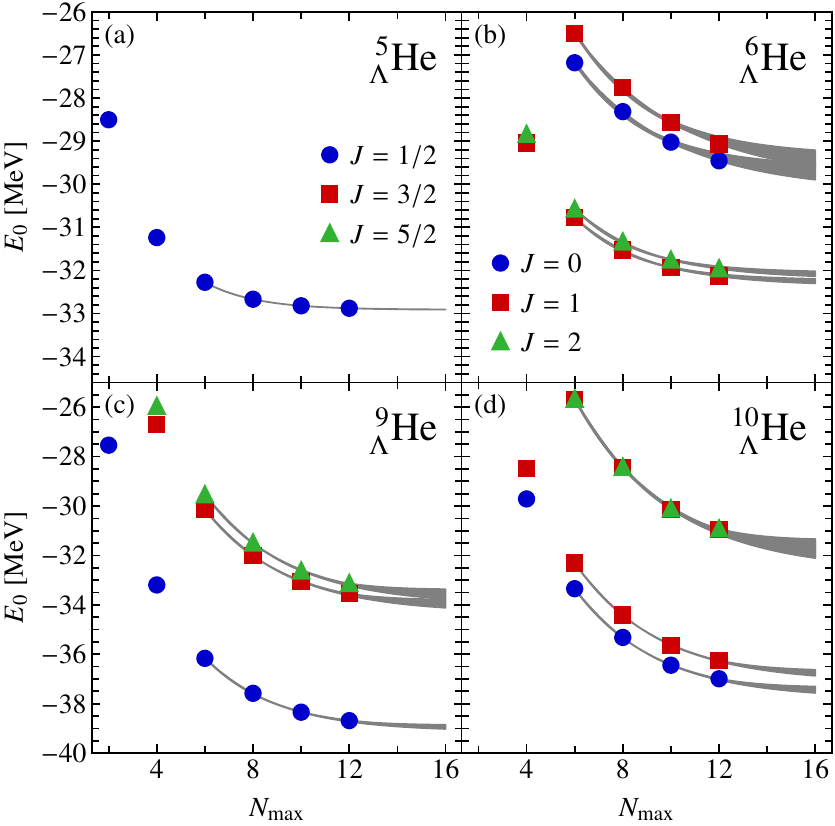}
  \caption{\label{fig:lhe convergence}%
    Absolute energies of the low-lying spectrum of four helium hypernuclei for the \YN{} interaction with $\Lambda_Y=\SI{700}{\MeVc}$ cutoff.
    The colors denote angular momenta: blue for $J=0$ ($1/2$), red for $J=1$ ($3/2$), green for $J=2$ ($5/2$) for even (odd) systems.
    The gray bands mark the envelope of the fit functions used to extrapolate the energies to infinite model-space size.
    Note that the nuclei shown in (a) and (c) are particle-stable while those in (b) and (d) are not, according to the calculation.
  }
\end{figure}

\begin{figure*}
  \centering
  \includegraphics{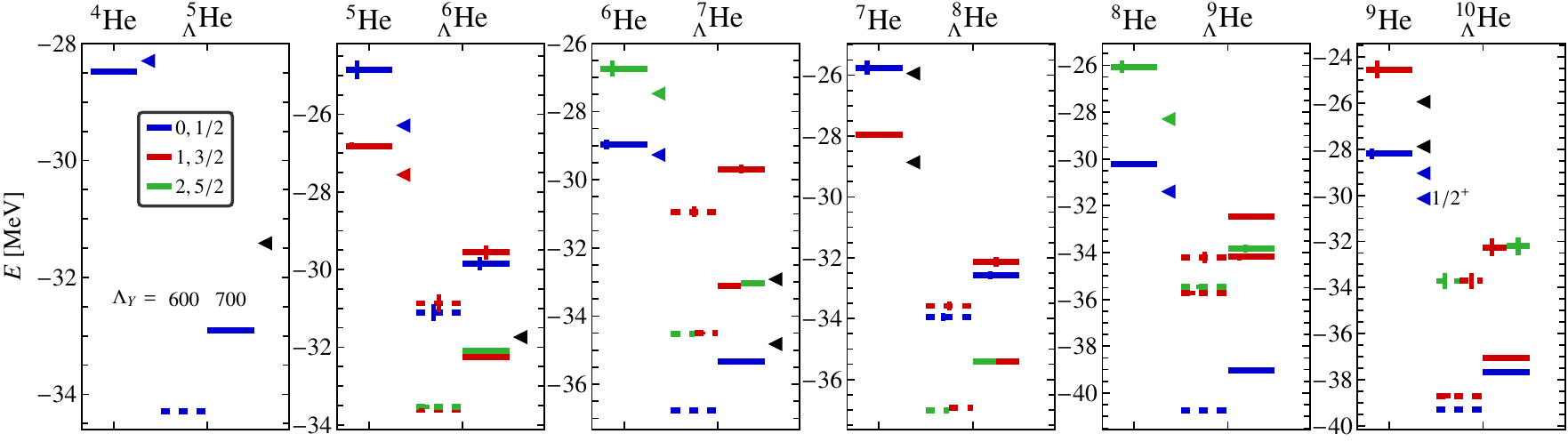}
  \caption{\label{fig:he chain}%
    Extrapolated energies of low-lying natural-parity states of hypernuclei along the helium chain.
    Shown are the nucleonic parents and the single-$\Lambda$ hypernuclei for two values of the \YN{} interaction regulator $\Lambda_Y=\SI{600}{\MeVc}$ (dashed lines) and $\Lambda_Y=\SI{700}{\MeVc}$ (solid lines).
    Experimental values \cite{Wang2012,Davis2005,Hashimoto2006,Tilley2002,Tilley2004,Gogami2016} are marked by triangles, vertical lines denote extrapolation uncertainties.
    The colors denote angular momenta: blue for $J=0$ ($1/2$), red for $J=1$ ($3/2$), and green for $J=2$ ($5/2$) for even (odd) systems.
    Unknown angular momenta are marked by black symbols.
  }
\end{figure*}

\begin{figure}
  \centering
  \includegraphics{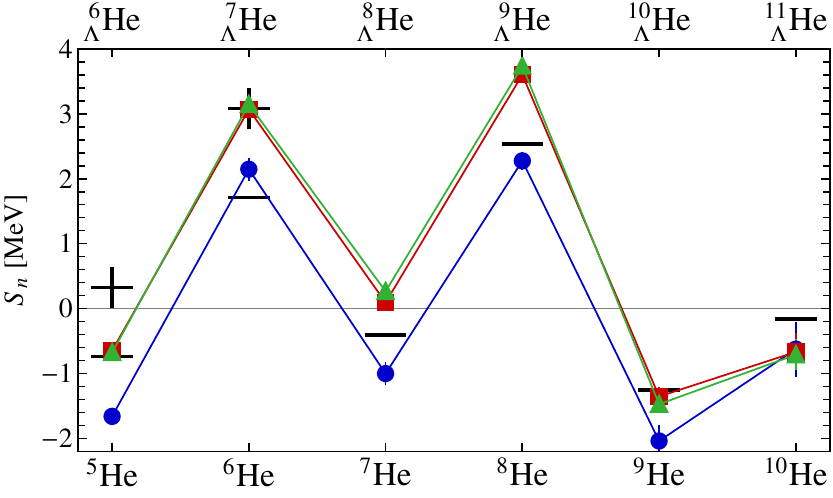}
  \caption{\label{fig:he sn}%
    Neutron separation energies of helium (hyper-)isotopes.
    Shown are the separation energies calculated for the nucleonic parents (blue circles) and for their daughter hypernuclei using the $\Lambda_Y=\SI{700}{\MeVc}$ (red squares) and $\Lambda_Y=\SI{600}{\MeVc}$ (green triangles) cutoffs.
    Experimental values are shown as black bars (crosses) for the (hyper-)nuclei.
    Vertical lines indicate extrapolation uncertainties.
  }
\end{figure}

\begin{figure*}%
  \centering%
  \includegraphics{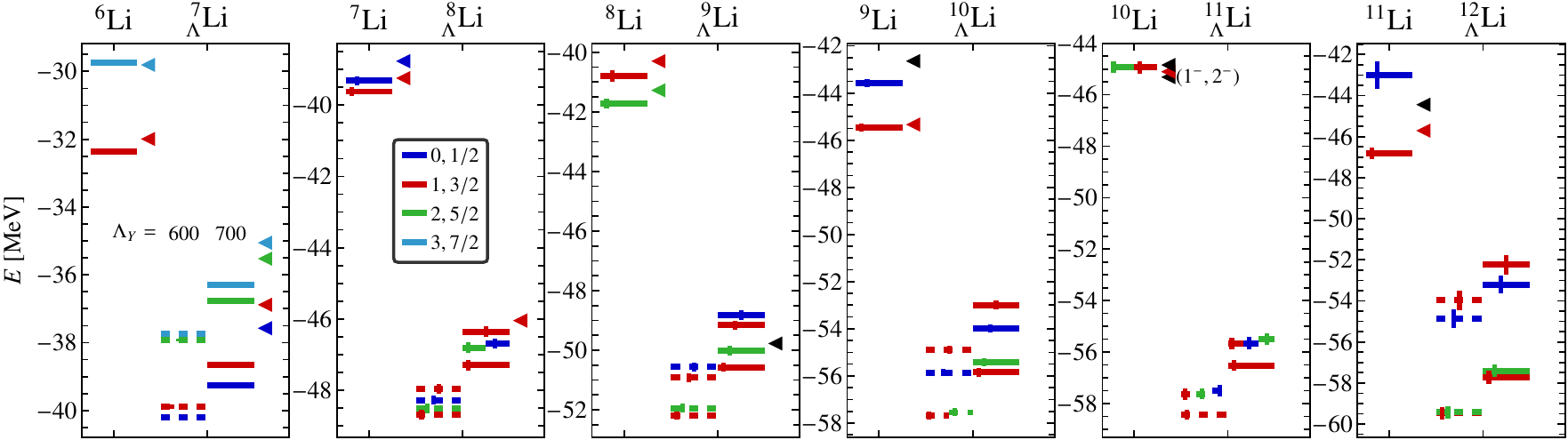}%
  \caption{\label{fig:li chain}
    Like \cref{fig:he chain}, but for the lithium chain.
    The colors denote angular momenta: blue for $J=0$ ($1/2$), red for $J=1$ ($3/2$), green for $J=2$ ($5/2$) and light blue for $J=3$ ($7/2$) for even (odd) systems.
    Unknown angular momenta are marked by black symbols.
    Experimental values are taken from \cite{Wang2012,Davis2005,Hashimoto2006,Tilley2002,Tilley2004,Kelley2012}.
  }
\end{figure*}

\begin{figure}
  \centering
  \includegraphics{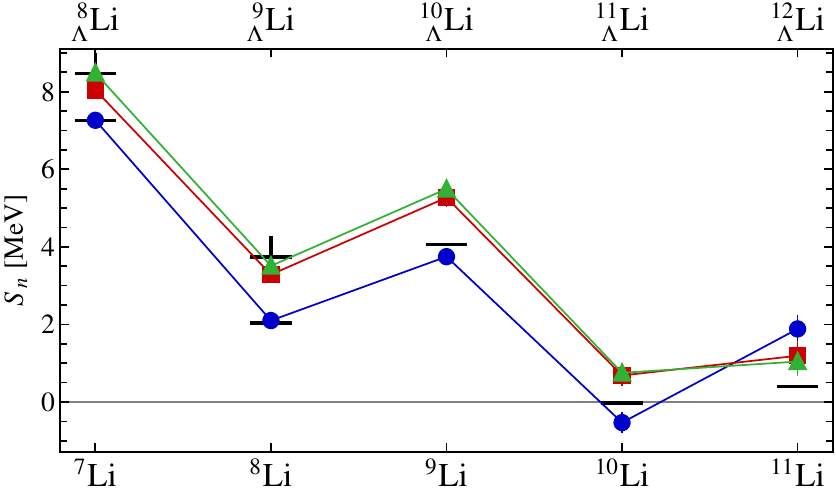}
  \caption{\label{fig:li sn}%
    Same as \cref{fig:he sn}, but for lithium (hyper-)isotopes.
  }
\end{figure}

We use the hypernuclear IT-NCSM to calculate the low-lying states of single\nobreakdash-$\Lambda$ hypernuclei throughout the helium and lithium isotopic chains.
The initial Hamiltonian consists of an \NN{} interaction at next-to-next-to-next-to-leading order (N\textsuperscript3LO) chiral effective field theory by \citet{Entem2003}, an \NNN{} interaction at N\textsuperscript2LO by \citet{Navratil2007}, and a \YN{} interaction at LO by \citet{Polinder2006}.
The regulators in the nucleonic sector are chosen as $\Lambda_N=\SI{500}{\MeVc}$.
For the hyperonic sector we employ two different cutoffs $\Lambda_Y=\SI{600}{\MeVc}$, $\SI{700}{\MeVc}$ in order to estimate the remaining uncertainty due to the truncation of the chiral expansion at leading order.
This Hamiltonian is SRG-evolved to a flow parameter $\alpha=\SI{0.08}{\fm\tothe4}$, as described in the previous section, and we apply the IT-NCSM to compute the four lowest states up to $\Nmax=12$ with a basis frequency of $\hbar\Omega=\SI{20}{\MeV}$.
We choose this basis frequency because it is close to the variational minimum for the $\Nmax$ range considered.
We always show the low-lying states with natural parity, which, in all the cases considered, is the parity of the calculated ground state.
Our calculations do not include continuum degrees of freedom, which are important for states close to threshold, and may lower the absolute energies of these states \cite{Papadimitriou2013,Baroni2013a,*Baroni2013b,Langhammer2015,Calci2016}.

The NLO terms of the \YN{} interaction have been derived by \citet{Haidenbauer2013}.
However, the 36 \YN{} scattering data points cannot constrain the 23 parameters (5 at LO, 18 at NLO) of the theory, especially the $p$-wave contact terms, so the authors additionally use $p$-wave \NN{} scattering phase shifts to fix them.
This introduces additional uncertainties, which is why we restrict ourselves to the LO interaction, which can be constrained by \YN{} data alone.

\Cref{fig:lhe convergence} shows $\Nmax$ sequences for a set of helium hypernuclei.
The ground state of \isotope[5][\Lambda]{He} (panel a) is practically converged at $\Nmax=12$.
The heavier system \isotope[9][\Lambda]{He} (panel c) converges more slowly so that we resort to extrapolation for the infinite-model-space result.
Having noticed that IT-NCSM calculations often show more Gaussian than exponential convergence for basis frequencies close to the variational minimum, we augment the simple exponential extrapolation used in, e.g., Refs.\ \cite{Maris2009,Roth2014}:
We fit three-parameter exponentials and a four-parameter extension with an additional $\Nmax^2$\nobreakdash-term to five different subsets of the $\Nmax$ sequence consisting of four to six points.
The mean and standard deviation of these ten fit results comprise the extrapolated value and its uncertainty.
The envelope of the extrapolation functions for the ground state shows only little spread and allows for a reliable extraction of the converged energy for \isotope[9][\Lambda]{He}.
The excited states, as well as the ground states of \isotope[6][\Lambda]{He} and \isotope[10][\Lambda]{He}, show a slower convergence.
These states are particle-unbound, which manifests itself in the different convergence behavior.

The uncertainty due to the choice of basis frequency is on par with those due to threshold and model-space extrapolation.
Using \isotope[7][\Lambda]{He} as a representative, we get extrapolated ground-state energies of \SI{-35.43(7)}{\MeV} for $\hbar\Omega=\SI{16}{\MeV}$, \SI{-35.33(6)}{\MeV} for $\hbar\Omega=\SI{20}{\MeV}$, and \SI{-35.17(10)}{\MeV} for $\hbar\Omega=\SI{24}{\MeV}$.

In \cref{fig:he chain}, we show extrapolated absolute energies of low-lying states of helium hypernuclei and their nucleonic parents.
The nucleonic calculation slightly underbinds the helium isotopes beyond \isotope[4]{He}, but correctly reproduces the particle-insta\-bil\-i\-ty of \isotope[5]{He} and \isotope[7]{He}.
Experimental data on hyperon separation energies is only available for the isotopes up to \isotope[7][\Lambda]{He}.
The \SI{700}{\MeVc} cutoff strongly overbinds \isotope[5][\Lambda]{He}, which is a long-standing issue with \YN{} interactions that reproduce the binding energies of the $A=4$ system \cite{Nemura2002}.
The overbinding in \isotope[6][\Lambda]{He} and \isotope[7][\Lambda]{He} is only a few hundred \si{\keV}, but this is in part due to the nucleonic calculation underbinding the helium isotopes.
The \YN{} interaction with \SI{600}{\MeVc} cutoff overbinds all these isotopes by about \SI{2}{\MeV}.

The nonstrange helium isotopes show a marked odd-even staggering that renders the odd isotopes unstable against neutron emission.
The additional binding provided by the hyperon does not suffice to stabilize \isotope[6][\Lambda]{He}, which is again an artifact of the overbinding in \isotope[5][\Lambda]{He}.
The ground-state doublet of \isotope[8][\Lambda]{He} is predicted to be at threshold within extrapolation uncertainties for the \SI{700}{\MeVc} cutoff.
The \SI{600}{\MeVc} cutoff puts the ground state \SI{0.26(6)}{\MeV} below the $\isotope[7][\Lambda]{He}+n$ threshold.

The staggering is also reflected in the neutron separation energies shown in \cref{fig:he sn}.
Our results agree with experiment at the level of a few hundred \si{\keV}, only the separation energy in \isotope[5]{He} is too low because the $3/2^-$ resonance is predicted too high.
Conversely, the separation energy in the daughter hypernucleus \isotope[6][\Lambda]{He} is too low because the Hamiltonian overbinds \isotope[5][\Lambda]{He}.
Unlike the absolute ground-state energies, the neutron separation energies of the hypernuclei are remarkably robust against variation of the regulator cutoff of the \YN{} interaction.

The separation energies of the hypernuclei follow the trend of their nucleonic parents with a shift, as expected by the \SI{1}{\MeV}-per-ad\-ditional-nucleon increase of the hyperon separation energy.
This behavior holds up to \isotope[9][\Lambda]{He}, which has a neutron separation energy of approx.\ \SI{3.6}{\MeV}, compared to \SI{2.3}{\MeV} in \isotope[8]{He}.
Surprisingly, the neutron separation energy of the next hypernucleus along the chain, \isotope[10][\Lambda]{He}, is essentially the same as the experimental value for \isotope[9]{He} and well in the unbound region.
At the $N=8$ shell closure, \isotope[11][\Lambda]{He} shows a similar behavior.
The hyperon provides very little additional binding, if any, for these very neutron-rich systems and the neutron drip line is the same as for the nonstrange isotopes.
From a mean-field perspective this may be interpreted as the hyperon lowering the $\nu0p_{3/2}$ orbit by \SI{1}{\MeV} while leaving the energy of the $\nu0p_{1/2}$ unaffected.

The low-lying states of hypernuclei and their nucleonic parents along the lithium isotopic chain are shown in \cref{fig:li chain}.
Overall, the calculations for the nucleonic parents are well-converged for the lighter isotopes and we get agreement between the calculated and experimental binding energies to better than \SI{1}{\MeV}.
The notable exception is \isotope[10]{Li}, where we fail to reproduce the parity inversion and the lowest negative-parity state is predicted at an excitation energy of \SI{1.2(5)}{\MeV}.
For the heaviest isotope considered, \isotope[11]{Li}, convergence is not complete at $\Nmax=12$, and the ground state is slightly overbound.
The excited state is probably a resonance, which converges slowly in the IT-NCSM and, therefore, has larger extrapolation uncertainties.
Given the halo nature of systems like \isotope[9]{Li} and \isotope[11]{Li}, the level of agreement with experimental data is fortuitous.

The description of the hypernuclear states is similar in quality to the more symmetric hypernuclei that we considered previously \cite{Wirth2016}.
The \SI{700}{\MeVc} cutoff reproduces the spectrum of \isotope[7][\Lambda]{Li} and the known ground-state energies with a systematic overbinding of \num{1} to \SI{2}{\MeV}.
The \SI{600}{\MeVc} cutoff overbinds more strongly by \num{2} to \SI{3}{\MeV} and produces smaller splittings among the hypernuclear doublet states.

The ground-state energies show a common trend with both cutoffs: the addition of a neutron to \isotope[7][\Lambda]{Li} lowers the ground-state by approximately \SI{8}{\MeV} with \SI{7}{\MeV} originating from the additional binding of the nucleonic core.
The remainder stems from the increase of the $\Lambda$ binding energy, which is in line with the commonly-observed value of \SI{1}{\MeV} per additional nucleon \cite{Davis2005}.
After the initial drop the ground-state energies continue to decrease more slowly with a slight odd-even staggering, following the trend of the nucleonic parents.

At \isotope[10][\Lambda]{Li}, the energies start to saturate, indicating proximity to the neutron drip line.
The core of \isotope[11][\Lambda]{Li}, which is predicted to be particle unstable with respect to neutron emission, is stabilized by the presence of the hyperon (cf.~\cref{fig:li sn}).
Note that the doublets originating from the $1^+$ ground state and the very low-lying $2^+$ excitation in \isotope[10]{Li} completely overlap, forming an isolated $3/2^+$ ground state and a nearly-degenerate triplet very close to the $\isotope[10][\Lambda]{Li}+n$ threshold.

The nucleon separation energies, shown in \cref{fig:li sn}, are less sensitive to the \YN{} cutoff than the absolute binding energies.
The nucleonic Hamiltonian reproduces the experimental values to a few hundred \si{\keV}, except for \isotope[11]{Li}, for which the separation energy is \SI{1.5(4)}{\MeV} too high.
As for the helium chain, the neutron separation energies of the hypernuclei are shifted to higher values compared to their nucleonic parents.
The \YN{} interaction with  \SI{600}{\MeVc} cutoff reproduces the experimentally known neutron separation energies of \isotope[8][\Lambda]{Li} and \isotope[9][\Lambda]{Li} almost within extrapolation uncertainties.
The larger cutoff provides systematically smaller separation energies.

While the ground-state doublet of \isotope[12][\Lambda]{Li} is particle-stable, the behavior of the neutron separation energies is different from the lighter isotopes:
the nucleonic core has a neutron separation energy of \SI{1.9(4)}{\MeV}, but the additional hyperon lowers this value to \SI{1.2(4)}{\MeV} (\SI{1.0(4)}{\MeV}) for the \SI{700}{\MeVc} (\SI{600}{\MeVc}) cutoff.
The experimental value for the \isotope[11]{Li} neutron separation energy is only \SI{0.40}{\MeV} and the calculation overestimates this value because the nucleonic Hamiltonian overbinds the \isotope[11]{Li} ground state.
Thus, when using a Hamiltonian that correctly reproduces the \isotope[11]{Li} ground state, the neutron separation energy of \isotope[12][\Lambda]{Li} will be lower and very close to the $\isotope[11][\Lambda]{Li}+n$ threshold.
The lack of additional binding due to the hyperon indicates that no neutrons beyond the $N=8$ shell closure will be bound.
The hypernuclear drip line is thus not different from the nucleonic one.

\section{Conclusions}
We calculate neutron-rich hypernuclei throughout the helium and lithium isotopic chains in an IT-NCSM framework using a Hamiltonian from chiral effective field theory.
For all but the lightest isotopes considered, this is the first time these hypernuclei have been addressed in an \emph{ab initio} framework with chiral interactions.
Our calculations for the ground and first-excited states of the nucleonic parents show good agreement with experimental data, except for \isotope[8]{He} and \isotope[11]{Li}, which show larger discrepancies, as well as \isotope[9]{He} and \isotope[10]{Li}, where the calculation fails to reproduce the parity inversion of the ground state.
The experimental spin-parity assignment of the \isotope[9]{He} ground state is currently under debate; see, e.g., Ref.\ \cite{AlKalanee2013}.
Some of these deficiencies can be attributed to missing continuum degrees of freedom from the calculation \cite{Papadimitriou2013,Baroni2013a,*Baroni2013b,Langhammer2015,Calci2016}.

Consistent with our previous findings in more symmetric hypernuclei, the absolute energies show a large cutoff dependence.
The \YN{} interaction with \SI{600}{\MeVc} cutoff overbinds systematically and the \SI{700}{\MeVc} cutoff is consistently closer to experiment.
If one takes a slightly more phenomenological approach with the aim of providing a good description of the available data, these results can be used to select a specific cutoff and tune the interaction parameters to achieve this.

The overbinding is relatively constant across the isotopic chains so that differential quantities like neutron separation energies are less sensitive to the cutoff of the \YN{} interaction regulator.
We achieve a reproduction of experimental neutron separation energies to better than \SI{100}{\keV} for the hypernuclei we considered, except for \isotope[6][\Lambda]{He}, for which the separation energy is skewed by overbinding of the \isotope[5][\Lambda]{He} ground state.
For the nonstrange nuclei, experimental values are reproduced to better than \SI{1}{\MeV}.
We find indications that \isotope[12][\Lambda]{Li}, which has an $N=8$ neutron shell closure, is at the drip line.
Contrary to the naive expectation, in the helium chain the hyperon does not provide additional binding to neutrons beyond $N=6$ so that the heaviest particle-stable isotope is \isotope[9][\Lambda]{He}.

\section*{Acknowledgments}
We gratefully acknowledge support by the BMBF through contracts 05P15RDFN1 (NuSTAR.DA) and 05P2015 (NuSTAR R\&D), the Deutsche Forschungsgemeinschaft through contract SFB 1245, and the Helmholtz International Center for FAIR.
Calculations for this research were conducted on the LICHTENBERG high-performance computer of TU Darmstadt and on the supercomputer JURECA \cite{JURECA} at Forschungszentrum Jülich.

\vspace*{2\baselineskip}
\newcommand{\doibase}[0]{https://doi.org/}%
\newcommand{\Eprint}[0]{\hfil\penalty100\hfilneg\href }
\bibliographystyle{apsrev4-1}
\bibliography{abbrev.bib,lightneutronrich.bib,extra.bib}

\end{document}